# Grammatical Parameters from a Gene-like Code to Self-Organizing Attractors[1]

Giuseppe Longobardi (Language and Linguistic Science, University of York) and
Alessandro Treves (Cognitive Neuroscience, SISSA, Trieste)

**Abstract**
Parametric approaches to grammatical diversity range from Chomsky's 1981 classical Principles & Parameters model to minimalist reinterpretations: in some proposals of the latter framework, parameters need not be an extensional list given at the initial state S0 of the mind, but can be constructed through a bio-program in the course of language development. In this contribution we pursue this lead and discuss initial data and ideas relevant for the elaboration of three sets of questions:
1) how can binary parameters be conceivably implemented in cortical and subcortical circuitry in the human brain?
2) how can parameter mutations be taken to occur?
3) given the distribution of parameter values across languages and their implications, can multi-parental models of language phylogenies, departing from ultrametricity, also account for some of the available evidence?

0.   Introduction

Parametric approaches to grammatical diversity range from Chomsky's 1981 classical Principles & Parameters model (especially see Fodor 1998, where this is suggestively termed 'the twenty questions' model) to minimalist reinterpretations: in some proposals of the latter framework, parameters need not be an extensional list given at the initial state $S_0$ of the mind, but can be constructed through a bio-program in the course of language development. In either type of model, anyway, the internal structure of a natural language is specified by the values taken by a set of binary variables, the parameters.

As the framework has gradually evolved over the last few decades, the number of parameters has grown remarkably, from e.g. the 10-15 envisaged in Baker (2001) to the hundreds (maybe even more) understood today to be necessary to fully describe human syntactic variability (Guardiano and Longobardi 2017); and their internal dependencies or implications have become increasingly structured. Despite the implications measurably restricting the space of syntactically possible languages (Bortolussi et al. 2012), it is clear that the $\approx 2^{12}$ natural languages existing today span only a sparse (and probably accidental) subset of all possible ones; yet the variability they encompass is remarkable. Nonetheless native speakers are able to acquire the particular parameter specification of their language effortlessly, and making use of the same neural machinery that all humans possess, which in turn is quite similar to at least that of other mammals. Recent work has revived the interest in the possibility of using parameters to explain the acquisition of these complex systems of variable knowledge and of locating them in the ontogenesis of grammars (e.g. Karimi and Piattelli Palmarini 2017, Biberauer 2019, Crisma et al 2020).

The conceptualization of parameter specification as an analogue of a genetic code, one that is transmissible across generations except for mutations (binary switches), has also afforded the reconstruction of phylogenetic trees of languages (Ceolin et al 2020) extending those produced by classical etymological methods, and comparable to the trees extracted from the actual genomic analysis of corresponding speakers' populations (Longobardi et al. 2015, Santos et al 2020).

---

[1] Contribution submitted to *A Cartesian dream: A geometrical account of syntax. In honor of Andrea Moro*. M. Greco and D. Mocci, eds., Rivista di Grammatica Generativa/Research in Generative Grammar. Lingbuzz Press ISSN 2531-5935.

Among the many paths that may lead further beyond the current boundaries of this analysis of grammatical variation as a cultural counterpart of a genetic code, we want here to stress three:

1) how can binary parameters be conceivably implemented in cortical and subcortical circuitry in the human brain? And how can their acquisition be represented in this context? In particular, we ask to what extent can Hebbian associative plasticity, asymmetric in time on a tens of msec scale, lead to the self-organization of alternative attractor structures that realize, in the speakers' mind, the set(s) of parameters of their native language(s).

2) how can parameter 'mutations' be taken to occur, as a possible effect of language contact, pressure from other linguistic levels, sheer chance or actually driven by evolutionary syntactic 'fitness'? Given the stability of established cortical attractor structures, reflected in fluent usage of a native language, under what conditions can they be destabilised by external or perhaps internal drives?

3) given the distribution of parameter values across natural languages and their implicational dependencies, can non-monoparental models of language phylogenies, departing from ultrametricity, also account for some of the available evidence? The development and the evaluation of alternatives to the mono-parental dendrograms of common cluster analyses might generate insights useful in the quantitative characterization of other evolutionary histories as well.

In the remaining three sections of this contribution we begin elaborating on these three sets of questions.

1. Cortical implementation

A major change in theoretical perspectives in neuroscience has resulted from the realization, imported from statistical physics, that some properties of the collective behaviour of many interacting elements are independent of the detailed properties of the individual elements. For example, phase transitions have been categorized into a number of universality classes, and a first-order phase transition is essentially the same animal, so to speak, whether it is water molecules that freeze or neurons that reactivate a memory by aligning to a partial cue. This realization has not had much of an impact in linguistics, however, largely because how language arises ontogenetically from the human brain and from the interactions among billions of neurons has often been outside the scope of purely linguistic research. The moment language production or acquisition is described in an information processing framework with a small number of dedicated variables, there is no collective behaviour and no statistics to talk about. Nevertheless, grounding linguistic phenomena at the neural level has the potential to clarify their development and their dynamics of change.

In the case of parameter values, it is plausible to think of them as alternative steady states of a history of synaptic weight changes, which has to be relatively short (the poverty of stimulus argument) and independent of explicit instructions. These two characteristics point to "Hebbian" associative synaptic plasticity. Formulated theoretically by Donald Hebb (1949), discovered experimentally fifty years ago and extensively investigated ever since, this dominant implementation of learning and memory in the brain sees a neuron A, which repeatedly participates in activating another neuron B, strengthen the efficacy of the synapse between them. Like any positive feedback system, it is self-reinforcing and can potentially be implemented in one shot, or at least very rapidly, based solely on positive evidence – that is, suitable examples. These are instances, even experienced passively, with no cause-effect relation, in which the activation of A is followed by the activation of B. Experiencing the examples leads to learning the 'rule' that B follows A, and the rule is expressed by the strengthened synapse that now makes A contribute more to activate B. It need not become a strict rule, obviously; it may get to be just a bias.

Jumping from the single neuron to the population level, billions of individual weight changes can 'dig up' *attractors*, the riverbeds where neural activity tends to flow. Models of associative memory networks have demonstrated how Hebbian plasticity can lead to multiple attractors of the *neural* dynamics, that is, neural activity changes at the tens-of-msec time scale (e.g. see Amit, 1989). With language parameters, we are interested in the possibility that it leads to such multiple attractors of neural dynamics but also, on the much longer scale of *synaptic* changes, to alternative sets of multiple attractors, which reflect the parameter values of a specific language. For example, a language that grammaticalises Number and marks it on nouns but does not spread it to adjectives (e.g., modern English) will endow newly acquired nouns with separate attractor states distinguished by Number, but not newly acquired adjectives; and the synaptic changes that enable these neural attractors are themselves stable and reliably reproduced each time (although only in a statistical sense), so they are attractors of synaptic dynamics.

How should we understand the self-organizing weight changes that lead native speakers of a language to acquire a specific set of parameters values? In general terms, as a sequence of phase transitions, each leading to the setting of a given parameter. Clues as to the nature of these transitions come from considering default parameter values. According to Crisma et al (to appear), three types of parameters can be distinguished based on the type of evidence necessary to set their values. In type 1a, positive evidence is available for both values of the parameter, but is more prominent and direct for one, say the '+' value, because the latter is associated to the independent need to learn some specific morphology. For example, encountering a plural adjective immediately shows that the Number feature spreads to adjectives, although also the converse situation, exposure to the same adjectival form both in contexts which are clearly singular and in others that are clearly plural would in principle, perhaps tentatively, lead the listener to set the '−' value. In type 1b, instead, exemplified by parameters linearly ordering nouns (or verbs) before vs. after their complements, the strength of the available evidence for either value is apparently the same. Finally, in Type 2 parameters positive evidence is only available for the + value, hence the − value can be naturally assumed to be the default one – for example, the parameter that establishes whether a head noun can take more than two arguments inflected with genitive case.

It seems straightforward to conceive of Type 2 parameters as emerging from a first-order phase transition – that is, if they switch from the default − value to the + value (Fig.1, left). Potential attractor states are associated with the minima of a free energy function, summarizing the interactions among the dynamical variables – the impulse rates of individual neurons collectively expressing linguistic constructs. Positive evidence induces plasticity that structures those interactions and thus "lowers" the local minimum of the free energy landscape, eventually turning it into a global attractor, while the default attractor does not change its absolute plausibility, it only loses out in relative terms. Something similar may occur with Type 1a parameters, where some evidence for the − value can be extracted from the data, but the dominant effect in the switch is still the lowering of the + attractor, if the specific language provides direct evidence in that sense (Fig.1, centre). The case of Type 1b parameters is most interesting, in that a default value seems to be ill-defined, and evidence for either − or + can be obtained with equal ease. This situation might be conceived as a very early second-order phase transition, in which two new local minima appear on the two sides of a primordial, linguistically unspecified *Ur-state*. Particularly in relation to word-order parameters, it is tempting to associate their setting to temporally asymmetric Hebbian plasticity (usually referred to as Spike Timing-Dependent Plasticity, Dan and Poo, 2004) which operates on a tens-of-msec scale, and spontaneously reinforces the tendency of e.g. one word category to occur before another, without any explicit instruction or rule learning. The self-organization of attractor structures may thus naturally account for apparently challenging phenomena, such as the setting of parameters with no obvious default value.

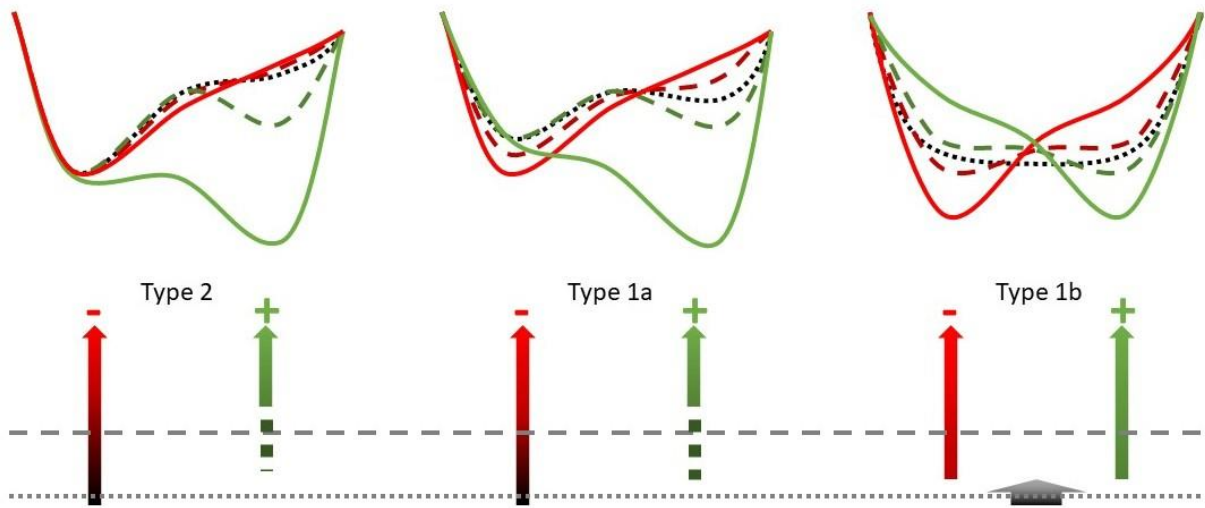

**Figure 1**. Varieties of phase transitions that may correspond to setting language parameters of different types. The top row sketches the "free-energy landscapes" at different stages of acquiring (or rather retaining) the + (green) and − (red) values of distinct types of parameters. The acquisition process runs upward in the bottom row. See the text, and Gafos and Benus (2006) as well as Dresher (2009) for the application of similar notions to phonological cognition.

2. Brain structures and parameter stability/resetting

Viewing an acquired full set of syntactic parameters as a specific *phase* in the operating regime of the cortical network of adult language speakers (comprised of billions of neurons) leads us to consider analysing the dynamics of another, much smaller network. This is the network of the syntactic parameters themselves, taken to be binary variables with an additional "zero" (i.e. 0) state (when a specific parameter need not be defined, given the values of others). Parameters interact through their (absolute) implications, which e.g. prevent parameter Y from taking value + if parameter X has value -, but also, potentially, through less studied graded implications, which make it only less likely for parameter Y to take value +, less likely but still possible. Phase transitions resulting from parameter switches in the cortical network correspond to ordinary parameter dynamics in the small parameter network, that would unfold over the years of language acquisition by the infant, under the influence of the prevailing language statistics (this contrasts with neural dynamics in the cortical network, which unfold much faster and throughout speaker's lifetime, during language comprehension and production). Given the hypothesis of multiple stable states for parameter dynamics, expressing the diversity of syntax across languages, how can we analyse it with a model?

A guiding principle can be that such an analysis should account for the relative stability of parameter values acquired with one's native language: for, they do change diachronically but are not constantly reset by transient exposure of individuals and generations to language samples with a different syntax (also see section 3 below). It is possible to incorporate a radical version of this principle by focusing on the concept of *ground states* of the network of interacting parameters. In this view, there must exist interactions among the parameters that help keep their values virtually "frozen", or constant, at least if the system is isolated. If one were to conceive of a fitness value assigned to every possible combination of parameter values, such ground states would have to be only locally optimal, in terms of such fitness function, otherwise one could not account for the long-term diversity and irreducible complexity of human languages. Therefore, any steady combination of parameter values would presumably contain some amount of "frustration", understood here as the inability to satisfy simultaneously all the hard and soft constraints expressed by the interactions: this would have the effect, among other things, of preventing languages from all becoming "simple" in the same way in the course of time. Anyway, when language input arrives, if incongruent, it can transiently perturb parameter values, but they tend to return to their ground state once the perturbation subsides.

The overarching question is, then, whether the parameter network admits (with all the necessary caveats of being a small system, with a few hundred variables only) *multiple* ground states. If so, parameter acquisition can be depicted as the dynamical evolution towards one of the many possible sets of ground states. It can be relaxational dynamics or driven dynamics: this depends on the relative strength of external (language) inputs vs internal constraints among parameters, but with enough of the latter to produce sufficient stability. Parameters can be conceptualized as specific instantiations of Potts variables, which is simply a historical term in physics (Domb and Potts, 1951) for variables that can take $m$ multiple categorical states (Siva et al., 2017): in our case a possible idealisation might be $m=2$: + and -; but also $m=3$: +, - and undefined or 0. In the parametric system here, + and – are taken to symbolise the standard oppositional values of binary parameters, while in fact 0 stands for a parameter state which is totally predictable by universal grammatical principles from the states of other parameters: they can be manifested in the language by expressions similar to those generated either by + or by -, according to each specific implicational rule. Notice that these are linguists' conventional notations, ultimately going back, in approaches like Biberauer (2019), Crisma et al. (2020) and Crisma et al. to appear, to the combination of two binary 'mental' operations: choice of presence/absence of the addition of a rule to the previous state of the mind, and presence/absence of predictability of the latter operation.

The ground states of a Potts network, a network obviously distinct from, but perhaps a useful model for the one we need to consider in the case of parameters, have been studied in statistical physics for a long time, and early results (Elderfield and Sherrington, 1983) point to a complex scenario, with different properties, particularly with reference to the multiplicity of ground states, when $m=2$, $m=3$, $m=4$ or $m>4$. A recent study (Ryom and Treves, 2023) shows that, while random networks entirely made up of Potts units with large $m$ are much slower than those comprised only of units with smaller $m$, in networks including both types of units the large-$m$ ones accelerate their dynamics and can become faster than the small-$m$ counterparts – a *speed inversion effect*, which might have wide implications. Clearly, the exact results are not directly transferable to the case of the parameter network, but their complexity reinforces the motivation to model it with a fully defined statistical physics formalism. In this sense, the crucial step is the development of an approximate description of the implications among parameters, where individual implications need not be reproduced, but their statistical structure should be preserved.

## 3. Parametric phylogenies and historical-evolutionary models

What relations does one expect among natural languages with respect to their parametric diversity? The problem has been first raised at the beginning of the century, most explicitly in Longobardi and Guardiano (2009), suggesting, against two centuries' scepticism, that syntax encodes a signal of language history.

If languages' internal structure is well captured by their parameter values, and the evolution in time of the latter is akin to relaxational dynamics in a Potts network, the analysis of the ground states of the statistically defined parameter network may reveal that they display properties analogous to those of the ground states of spin glasses (Mézard et al, 1987), as Potts networks under certain conditions behave like spin glasses. There is, however, much potential complexity that needs to be elucidated. Spin glass ground states, for example, have been hypothesized to comprise an *ultrametric* set, which would nicely match the simple mono-parental tree structure that is commonly assumed to describe phylogenetic relations across languages, and hence across sets of syntactic parameters. Ultrametric sets can in fact be viewed as generated by a simple mono-parental tree, in which 3 nodes of the last generation are all at the same distance from each other if they all share the most recent common ancestor; and are at the vertices of an isosceles triangle with long sides, if only 2 of them do so. In an ultrametric set, then, no (terminal) node is intermediate between two other nodes. It is not clear, though, to what extent ultrametricity is a property only of the mathematical solution or, indeed, of spin glasses themselves, and/or whether it applies to real examples of such materials, rather than to their mathematically defined models.

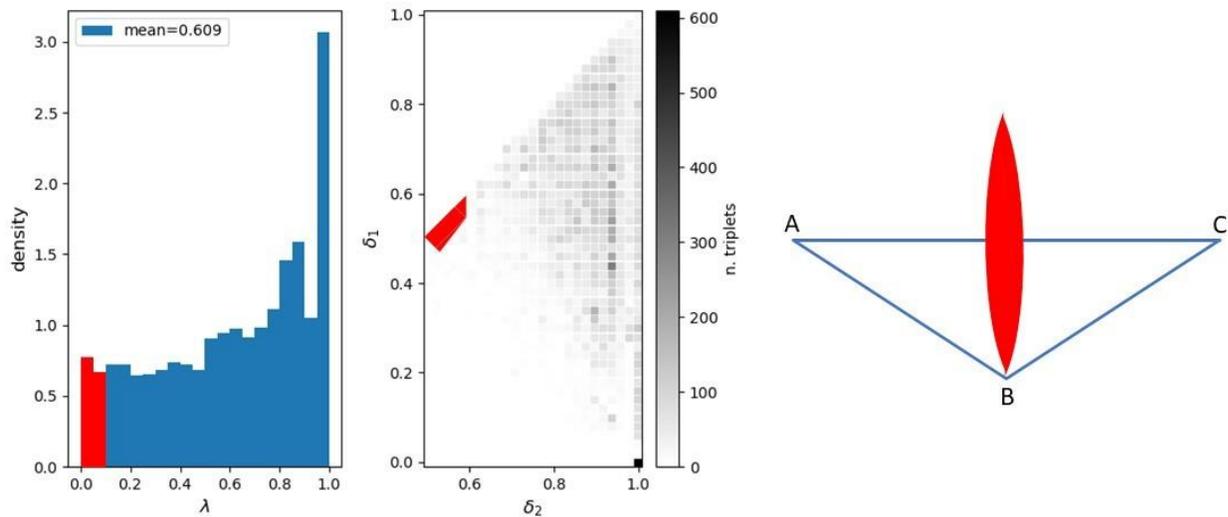

**Figure 2.** An analysis of ultrametricity may proceed by considering the mutual distances {D} (calculated as in Ceolin et al. 2021, cf. Supplementary Material there) among triplets of elements (right; in our case the elements are sets of parameter values, i.e., languages); the scatterplot of the two ratios $\delta_1 = D_{min}/D_{max}$ and $\delta_2 = D_{med}/D_{max}$ span the range described by the triangle in the middle; and an ultrametric index $\lambda$ can be defined as $\ln(\delta_1/\delta_2)/\ln(\delta_1 \ast \delta_2)$ (Treves 1997). Ultrametric triplets are characterized by $\delta_2 = \lambda = 1$ (the rightmost bin in the histogram on the left) while the least ultrametric ones are those where $\delta_1 \approx \delta_2 \approx 0.5$, so that $\lambda \approx 0$, e.g. those falling in the red region of the three panels. See text.

A simple phylogenetic tree would display ultrametricity among its terminal nodes if their divergence from common ancestors is produced by a "smooth" flow through many random parameter switches, such that on average each of the descendants comes to be at a similar distance from the progenitor, but in a random direction – and hence at similar distances from each other. If, instead, a terminal node comes to be exactly in between two other nodes, this indicates that it may have not moved much from the ancestor, hence it is in some sense more "archaic" than its peers. Interestingly, in the large dataset collected in Crisma et al (2020: cf. Supplementary Information) and analysed phylogenetically in Ceolin et al (2021) there are several examples of languages whose syntax, expressed as a simple vector of parameter values and given a straightforward definition of distance, lies in between that of other seemingly unrelated languages. Table 1 below gives three examples of such triplets with the corresponding distance values. There appear to be fewer triplets in a similar situation among languages with a common ancestor.

The occurrence of such triplets, again, would appear to be unlikely in a simple model of protracted random "genetic" drift, in which one expects that after a sufficiently long time all terminals have drifted at approximately the same distance from each other.

| Triplet 1 | | | Triplet 2 | | | Triplet 3 | | |
|---|---|---|---|---|---|---|---|---|
| ⌐0.375- | Hindi (Marathi) | -0.381⌐ | ⌐0.389- | Udmurt | -0.381⌐ | ⌐0.350- | Pashto | -0.364⌐ |
| Mandarin, Cantonese | --0.750-- | Arabic, Hebrew | Western Basque | --0.667-- | Malagasy | Tamil, Telegu | --0.700-- | Japanese |

**Table 1.** Three example triplets of unrelated languages that violate ultrametricity, with language distance values.

From a preliminary analysis, the most extreme violations of ultrametricity – the "forbidden regions" marked red in Fig.2 – appear indeed to concern languages belonging to families whose syntax has not revealed traceable signals of common ancestry in the statistical study of Ceolin et al (2021), whose final tree of hypothesised language clusters (adapted in the colours) is reported in Fig. 3.

While a statistical analysis of the entire sample of 58*57*56/6=30,856 triplets will be presented elsewhere, one can readily observe that for many of the languages in the sample the notion of a prolonged drift is not that appropriate, as they appear to have diverged in relatively recent times, and to have undergone just a few parameter changes. If part of the observed language diversity stems from the dominant contribution of a few changes, how do the effective interactions between specific pairs of parameters contribute to the overall statistics?

Consider two parameters A and B, that do take distinct values and hence produce diversity in the sample. If there is a strong interaction or implication between them, for example B is defined only if A takes the non-default value +, one should observe languages –A,0B, +A,-B and +A,+B, so that this parameter pair would contribute to the ultrametric character of the overall statistics. If instead A and B are unrelated, effectively non-interacting, one should observe languages –A,-B, -A,+B, +A,-B and +A,+B, in which each of the 4 types "sits" at the corners of a rectangle, as it were, and is therefore intermediate, in terms of distances, between the two languages along the adjacent corners, thus contributing to break ultrametricity. In tracing evolutionary history, the observation of all 4 typologies might be interpreted in some cases as deriving from multi-parental heritage: for example, a mother tongue with the "+A gene" has combined with a "father tongue" with the +B gene. Alternatively, the switch from, say, -B to +B may have occurred independently of the one from a- to +A, in both -A and +A lineages (what can be termed "convergent evolution"), or even multiple times, back and forth. This latter scenario is plausible particularly for parameters that do not affect the effectiveness or viability of language as a communication system – perhaps the vast majority of those observed to change. It may be noted that in this regime optimality arguments would then be ill-suited to understand parametric variation.

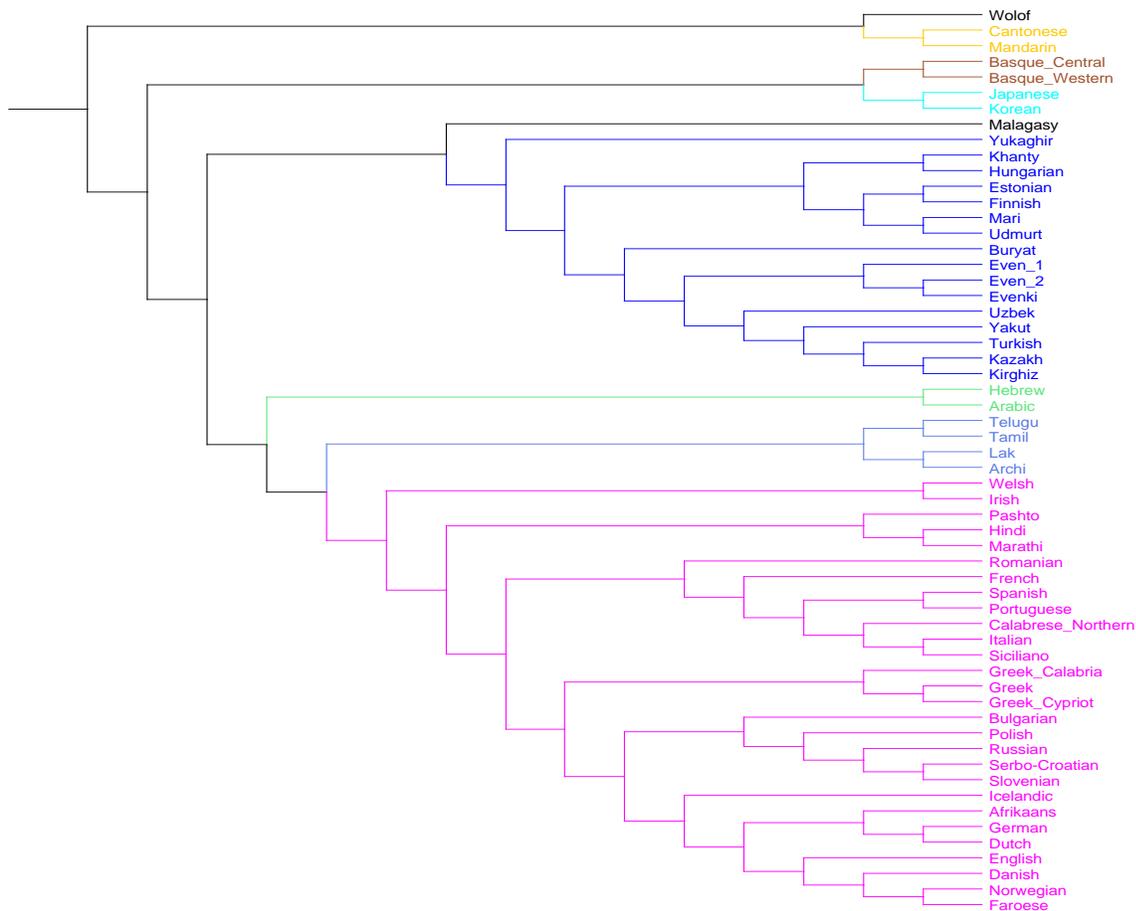

**Figure 3** - A UPGMA tree from all the syntactic distances: colours represent groups identified as plausible by the statistical test in Ceolin et al. (2021).

The availability of a statistical database of 58 languages, described by 94 parameters, therefore facilitates the data-driven (as opposed to theory-driven) inference of implications, or parameter interactions, which can proceed with the methods developed for network reconstruction (Roudi and Hertz, 2011), ultimately helping to assess mechanisms of language change. Do examples of ultrametricity violation reflect processes different from random drift, for example the so-called horizontal interactions with neighbouring descendants of other language ancestors? Or multiple parents, if the latter can be identified as a special case (perhaps in instances of creolisation)? Or maybe can these situations occur also within glassy dynamics, when parameters undergo rare changes in their values, so that the terminal nodes end up at very different distances from a putative common ancestor, at variance with the model of a smooth drift?

These questions can further motivate a detailed analysis of the distribution of syntactic parameter values that eventually will enable the formulation of statistical physics models, for example of the type attempted by DeGiuli (2019), that may help understand both language evolution, across populations of speakers, and language acquisition, within the brain of each speaker (Friedmann et al, 2021).

**Acknowledgments**. We are grateful to an anonymous reviewer and for very fruitful discussions to Andrea Ceolin and Kwang Il Ryom.